\documentclass[%
 reprint,
superscriptaddress,
 amsmath,amssymb,
 aps,
]{revtex4-1}

\usepackage{graphicx}% Include figure files
\usepackage{subcaption}%Includes subfigures
\usepackage{dcolumn}% Align table columns on decimal point
\usepackage{bm}% bold math
\usepackage{epstopdf}
\usepackage{upgreek}
\usepackage{graphicx}
\usepackage{hyperref}
\usepackage[justification=justified, format=plain]{caption} % 'format=plain' avoids hanging indentation

\begin{document}
%\preprint{APS/123-QED}

\title{Predominant Contribution of Direct Laser Acceleration to High-Energy Electron Spectra in a Low-Density Self-Modulated Laser Wakefield Accelerator}

\author{P. M. King}% 
\email{king100@llnl.gov}
\affiliation{NIF and Photon Sciences, Lawrence Livermore National Laboratory, Livermore, California 94550, USA}
\affiliation{Department of Physics, University of Texas at Austin, Austin, TX 78712, U.S.A.}
\author{K. Miller}
\affiliation{Department of Physics, University of California Los Angeles, Los Angeles, California, USA 90095}
\author{N. Lemos}
\affiliation{NIF and Photon Sciences, Lawrence Livermore National Laboratory, Livermore, California 94550, USA}%Lines break automatically or can be forced with \\
\author{J. L. Shaw}
\affiliation{Laboratory for Laser Energetics, University of Rochester, Rochester, NY, 14623 USA}
\author{B. F. Kraus}
\affiliation{Department of Astrophysical Sciences, Princeton University, Princeton, New Jersey 08544, USA}
\author{M. Thibodeau}
\affiliation{NIF and Photon Sciences, Lawrence Livermore National Laboratory, Livermore, California 94550, USA}
\author{B. M. Hegelich}
\affiliation{Department of Physics, University of Texas at Austin, Austin, TX 78712, U.S.A.}
\author{J. Hinojosa}
\affiliation{Center for Ultrafast Optical Science, University of Michigan, Ann Arbor, MI, 48109-2099, USA}
\author{P. Michel}
\affiliation{NIF and Photon Sciences, Lawrence Livermore National Laboratory, Livermore, California 94550, USA}
\author{C. Joshi}
\affiliation{Department of Electrical Engineering, University of California Los Angeles, Los Angeles, California, USA 90095}
\author{K. A. Marsh}
\affiliation{Department of Electrical Engineering, University of California Los Angeles, Los Angeles, California, USA 90095}
\author{W. Mori}
\affiliation{Department of Physics, University of California Los Angeles, Los Angeles, California, USA 90095}
\author{A. Pak}
\affiliation{NIF and Photon Sciences, Lawrence Livermore National Laboratory, Livermore, California 94550, USA}
\author{A. G. R. Thomas}
\affiliation{Center for Ultrafast Optical Science, University of Michigan, Ann Arbor, MI, 48109-2099, USA}
\author{F. Albert}
\affiliation{NIF and Photon Sciences, Lawrence Livermore National Laboratory, Livermore, California 94550, USA}

\date{\today}% It is always \today, today,

\begin{abstract}

    The two-temperature relativistic electron spectrum from a low-density ($3\times10^{17}$~cm$^{-3}$) self-modulated laser wakefield accelerator (SM-LWFA) is observed to transition between temperatures of $19\pm0.65$ and $46\pm2.45$~MeV at an electron energy of about 100~MeV. When the electrons are dispersed orthogonally to the laser polarization, their spectrum above 60~MeV shows a forking structure characteristic of direct laser acceleration (DLA). Both the two-temperature distribution and the forking structure are reproduced in a quasi-3D \textsc{Osiris} simulation of the interaction of the 1-ps, moderate-amplitude ($a_{0}=2.7$) laser pulse with the low-density plasma.  Particle tracking shows that while the SM-LWFA mechanism dominates below 40~MeV, the highest-energy ($>60$~MeV) electrons gain most of their energy through DLA. By separating the simulated electric fields into modes, the DLA-dominated electrons are shown to lose significant energy to the longitudinal laser field from the tight focusing geometry, resulting in a more accurate measure of net DLA energy gain than previously possible.

\end{abstract}

\maketitle

Laser wakefield acceleration (LWFA) produces beams of relativistic electrons~\cite{Tajima1979,Esarey2009} that can be used to generate directional x-rays~\cite{Corde2013} useful for imaging biological samples~\cite{Cole2015}, laser-driven shock fronts~\cite{Wood2018} and surface defects in alloys~\cite{Hussein2019}. These applications are enabled by the highly advantageous features of LWFA-driven x-rays~\cite{Albert_apps}: a small $\upmu$m-sized source, low divergence ($<50$~mrad), extremely broad (keV--MeV) range of photon energy, and synchronization with the drive laser to within 10s of femtoseconds. These attractive characteristics are attained by operating LWFA in the blowout regime~\cite{bib:WeiLu2007} with Joule-class, sub-50-fs-duration laser pulses.

By contrast, many High Energy Density Science (HEDS) facilities such as the National Ignition Facility at the Lawrence Livermore National Laboratory, the OMEGA Laser at the University of Rochester, the Z-Machine at Sandia National Laboratories, and the Laser Mégajoule at the Commissariat à l’Energie Atomique are all coupled to picosecond-duration, kilojoule-class laser systems. Understanding the relationships between temperature, pressure, and density in the extreme environments created at these large facilities is crucial and has implications in planetary science~\cite{Lee2006}, inertial confinement fusion~\cite{Lindl1995}, and laboratory astrophysics~\cite{Bulanov2009}. A limiting factor has been the quality of x-ray sources available to diagnose experiments. LWFA-driven sources have the potential to enable ultrafast resolution of dynamic experiments and provide measurements with unparalleled spatial resolution~\cite{Albert2017,NunoBrem,King2019}. While the promise of such x-ray sources is evident, optimization of LWFA-driven x-rays with the picosecond-duration lasers available at HEDS facilities necessitates a detailed understanding of the underlying physics of electron beam generation mechanisms---not in the LWFA regime, but in the self-modulated LWFA (SM-LWFA) regime~\cite{bib:modena_nat_1995}, which is as yet incomplete.

When the drive laser pulse satisfies $c\tau_{l} > \lambda_{p}$ (where $c$ is the speed of light, $\tau_{l}$ is the laser pulse duration and $\lambda_p = 3.3 \times 10^{10} (n_e)^{\frac{1}{2}} \,[\upmu$m$]$ is the plasma wavelength with electron density $n_e \,[$cm$^{-3}]$) and the plasma is under-dense ($n_{e} < n_{c}$, where $n_{c}$ is the density at which the laser frequency, $\omega_{0}$, equals the plasma frequency, $\omega_{p}$), the laser pulse can drive relativistically propagating plasma waves through the combined action of the self-modulation and Raman forward scattering instabilities~\cite{Joshi1981}. The plasma wave amplitude can become large enough to inject electrons~\cite{Dawson1959} into the plasma wave and accelerate them due to the longitudinal electric field of the plasma wave~\cite{bib:modena_nat_1995}. Electrons injected off the main laser axis undergo transverse oscillations due to the restoring forces in the plasma wave. These transverse oscillations can be amplified by the electric field of the overlapping picosecond laser pulse and converted into a longitudinal acceleration through the $\vec v \times \vec B$ force of the laser in a mechanism known as direct laser acceleration (DLA)~\cite{PukhovPoP1999,Gahn1999,Zhang2015,Zhang2016,Shaw2017}. In this situation, the relative contributions of SM-LWFA and DLA to the final electron energy remains poorly understood. 

Prior work on SM-LWFA has attributed the copious charge of high-energy electrons to self-trapping and breaking of the longitudinal plasma wave~\cite{Najmudin2003,Decker1994}. Electrons with energies larger than the dephasing-limited energy gain were also observed~\cite{Gordon1998,Malka2002}, but were always attributed to acceleration by the plasma wave. The role of DLA in LWFA was first suggested by Pukhov~\cite{PukhovPoP1999} and has been experimentally investigated only recently. In a quasi-blowout regime, the laser pulse was lengthened to overlap with a full plasma period, and electrons in the high-energy tail of the accelerated electron spectrum showed a fork-like splitting when dispersed perpendicular to the laser polarization direction~\cite{Shaw2014,Shaw2017}. This fork-like structure was attributed to DLA through PIC simulations, but the analysis did not include the contribution of the longitudinal field from the focused laser in the DLA process.. In a high-density ($\sim 10^{20}$~cm$^{-3}$), short-pulse (50~fs) SM-LWFA regime, the high-energy electron beam tail was experimentally attributed to DLA~\cite{Adachi2006}, but without any clear experimental signature. In a long-pulse (650~fs), high-intensity ($I = 3\times 10^{20}$~W/cm$^2$) regime, DLA was inferred from PIC simulations to be the main acceleration mechanism in an ion channel~\cite{Mangles2005}. The role of DLA in an SM-LWFA was anticipated in a recent experiment on developing a betatron-radiation-based x-ray source, but no direct experimental evidence for DLA was presented in that work~\cite{Albert2017,Lemos_2016}.

In this Letter we show experimental evidence, verified with quasi-3D particle-in-cell (PIC) simulations using \textsc{Osiris}~\cite{Fonseca2002,Davidson2015}, that DLA occurs concurrently with SM-LWFA but is the dominant contributor to the highest-energy electrons in the low-density $(\omega _p \ll \omega _0)$ regime of SM-LWFA, where the laser power---although it is greater than that needed for relativistic self-focusing---is insufficient to produce a totally evacuated ion channel inside the laser pulse. Our experimental work shows that, for a low plasma density ($\omega_0/\omega _p = 57$) and a 1-$\upmu$m, nominally 1-ps laser with moderate amplitude $a_0 = 8.5 \times 10^{-8} \lambda I^{1/2} \approx 2.2$, the accelerated electrons exhibit a two-temperature distribution. Here $a_0$ is the normalized vector potential and $\lambda$~[nm] is the laser wavelength. Full-scale quasi-3D PIC simulations confirm that the longitudinal field of the plasma wave excited by the SM-LWFA process mainly contributes to the low-temperature portion of the spectrum, whereas DLA is the dominant acceleration mechanism for the high-energy (temperature) electrons. When the electrons are dispersed orthogonally to the laser polarization direction, a fork-like structure~\cite{Shaw2017} characteristic of DLA is observed for electrons with energies above 60~MeV. This is the first direct experimental characterization, confirmed by quasi-3D PIC simulations, of DLA in an SM-LWFA in the picosecond, high-energy regime relevant to HEDS experiments. 

\begin{figure}[ht]
\centering
\includegraphics[trim={20mm 30mm 0mm 0mm}, clip=true, scale = .45]{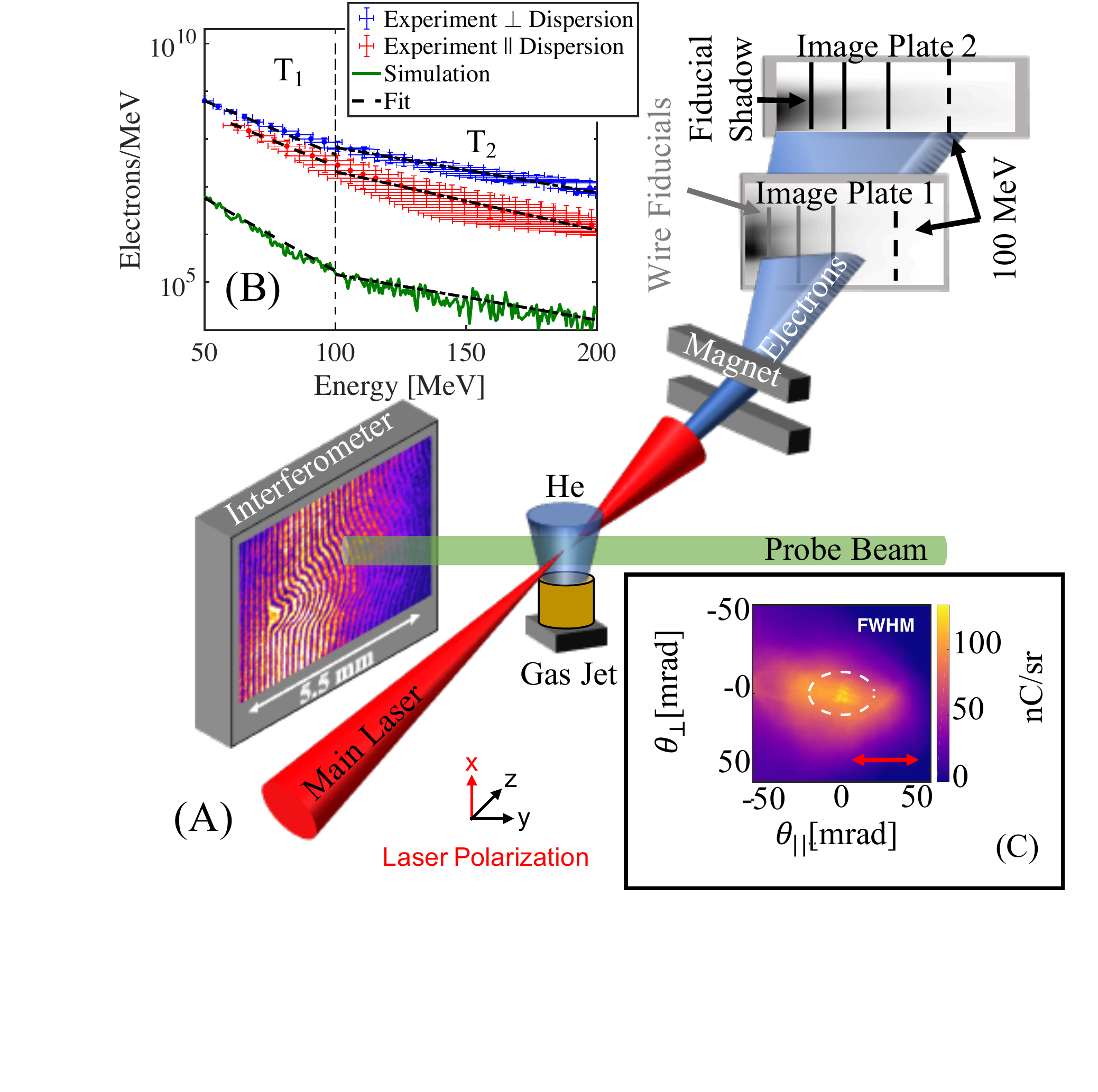}
\caption{(A)~Experimental setup. The electron beam is dispersed by a 0.6~T magnet onto two BAS-MS image plates after passing through three wire fiducials. A frequency-doubled probe beam is co-timed with the main pulse and provides on-shot interferometry of the plasma channel with a magnification of 3. (B)~The electron energy spectrum for two different shots using identical laser and plasma parameters dispersed perpendicular and parallel to the laser polarization in blue and red, respectively, along with the simulated electron spectrum in green. All three spectra are fit to single-temperature distributions below ($T_1$) and above ($T_2$) 100~MeV; the two regions are separated by a dashed black line. The experimental spectra exhibit shot-to-shot variations of the single-shot laser system. (C)~An undispersed electron beam profile.}
\label{fig:Fig1}
\end{figure}

We conducted the experiment on the Titan laser system at Lawrence Livermore National Laboratory (Fig.~\ref{fig:Fig1}). Titan, a $0.7^{+0.3}_{-0.1}$~ps, 120~J, Nd:Glass laser, was focused with an $f$/10 off-axis parabolic mirror into a 10-mm, supersonic He gas jet with electron density $n_e = 3 \times 10^{17}$~cm$^{-3}$ (measured using interferometry). This configuration created peak laser intensities reaching $ I = 6.4\times10^{18}$~W/cm$^{2}$, in a spot with 50$\%$ of the total energy contained in a 30-$\upmu$m radius. The ratio $P_\textrm{peak}/P_\textrm{crit} \approx 1.6$, where $P_\textrm{peak}$ is the peak laser power and $P_\textrm{crit} = 17 \times 10^{9} \left(\frac{\omega}{\omega_p}\right)^2 = 56$~TW is the critical power for relativistic self-focusing in the plasma~\cite{Esarey2009,bib:WeiLu2007}.

\begin{figure}[ht]
\centering
\includegraphics[trim={1mm 1mm 1mm 1mm}, clip=true, scale = .33]{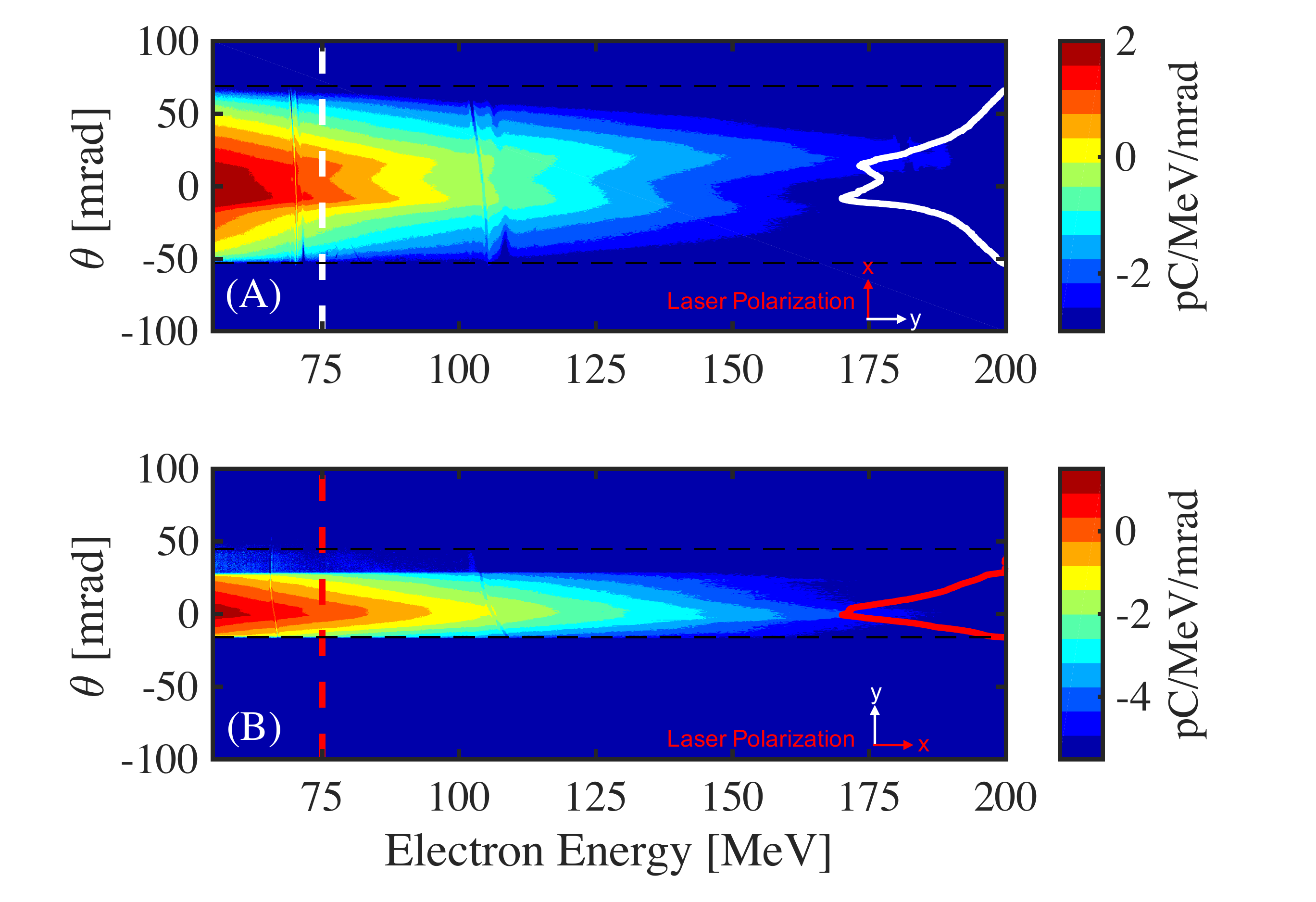}
\caption{Measured electron energy spectra for a plasma with electron density $n_e = 3\times10^{17}$~cm$^{-3}$ dispersed (A)~perpendicular and (B)~parallel to the linear laser polarization direction. The contrast is adjusted and a line-out along the dashed line is plotted (solid line) to emphasize the forking feature in the dispersed electron profile. %The signal above 100~MeV is multiplied by 10 and 20 in (A) and (B), respectively, to emphasize the spectrum at high energy.
The dashed black line indicates the acceptance aperture of the magnet. Note that (A) and (B) were taken on two different shots with similar laser energies.}
\label{fig:Fig2}
\end{figure}

The electron spectra shown in Fig.~\ref{fig:Fig2} dispersed (A)~perpendicular and (B)~parallel to the laser polarization were each fit using a single-temperature function ($A e^{-\frac{E}{T}}$, with amplitude $A$ and effective temperature $T$ in MeV) below and above 100~MeV [Fig.~\ref{fig:Fig1}(B)], yielding low and high temperatures of $T_1=19\pm0.65$~MeV and $T_2=46\pm2.45$~MeV, respectively. The perpendicularly dispersed electron signal in (A) (after it is converted to give a linear energy dispersion) shows a clear fork-like structure that begins at electron energies of $\sim$60~MeV. At 75~MeV, the FWHM of the divergence angle of this fork is 43~mrad [white curve in Fig.~\ref{fig:Fig2}(A)]. The mean total charge contained in this portion of the spectrum ($>60$~MeV) is $1.14\pm0.69$~nC. When the electrons are dispersed in the same plane as the laser electric field [Fig.~\ref{fig:Fig2}(B)], no forking structure is seen, and the FWHM beam divergence is instead 21~mrad [red curve in Fig.~\ref{fig:Fig2}(B)] at the same energy. The elliptical beam profile of the electrons shown in Fig.~\ref{fig:Fig1}(C) gives the overall full-angle divergence at half-maximum charge of the electron beam in the two planes as 47 and 27~mrad in the $x$ and $y$ directions, respectively, consistent with the dispersed spectra.

The forking structure gives clear evidence that electrons above 60~MeV are gaining some or most of their energy by the DLA process~\cite{Shaw2014,Shaw2017}. Electrons accelerated mainly through DLA generally exhibit higher energy and greater divergence along the laser polarization direction compared to electrons accelerated predominantly through SM-LWFA. This larger divergence is evident in the forking structure seen only for high-energy electrons dispersed perpendicular to the laser polarization, as in Fig.~~\ref{fig:Fig2}(A).

To discern the relative contribution of the various mechanisms to the final energy of the electrons, we simulated the full acceleration process with particle tracking using the quasi-3D algorithm of the \textsc{Osiris} PIC simulation framework~\cite{Fonseca2002,Davidson2015} for laser and plasma parameters similar to those used in the experiment (see supplemental material). This algorithm allows us to unambiguously determine the work done by the longitudinal field of the plasma wave $(E_{z,m=0})$, as well as the transverse $(E_{x,m=1})$ and longitudinal $(E_{z,m=1})$ fields of the laser. This has allowed us to more correctly determine the overall DLA contribution. Here $m=0$ and $m=1$ refer to the cylindrical modes corresponding predominantly to the wake and the laser, respectively. 

\begin{figure}[ht]
\centering
\includegraphics[trim={1mm 1mm 1mm 1mm}, clip=true, scale = .34]{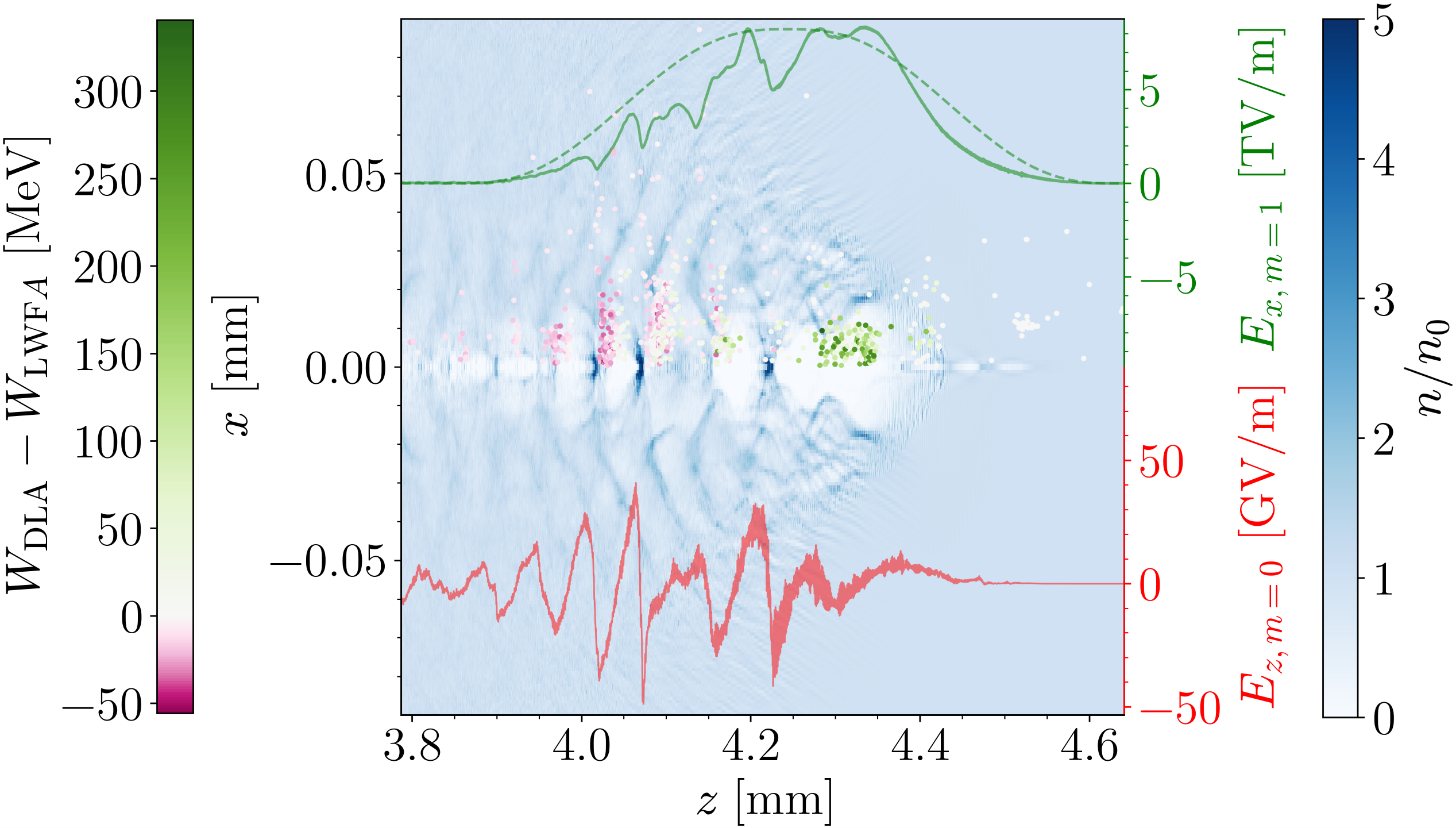}
\caption{Snapshot of the electron density profile after 4.64~mm of propagation (left to right) through the plasma; $z$ and $x$ are the longitudinal and transverse directions, respectively. Also shown are the $m=0$ longitudinal electric field (SM-LWFA) overlaid in red and the $m=1$ transverse electric field envelope (DLA) in green. The dashed green line shows the vacuum laser field envelope at the focus.  The tracked electrons, with $x$ positions given by their radial distance (only half-space is shown), indicate where in space each acceleration mechanism is dominant.  The charge density has been integrated in $\theta$.}
\label{fig:Fig3}
\end{figure}

Figure~\ref{fig:Fig3} shows the envelope of the transverse laser field $E_{x,m=1}$ (green), the plasma density (blue) and the on-axis longitudinal electric field of the plasma wave $E_{z,m=0}$ (red) 4.64~mm into the plasma. Clear modulation of both the laser envelope and plasma waves is evident.  However, a hydrodynamic channel is not fully formed (not shown) within the laser pulse; the ion density remains above $0.9n_0$ across the first bucket  (potential well) and above $0.7n_0$ where the laser field is of significant amplitude, where $n_0$ is the initial plasma density.
The wavelength of the plasma wave is increased for the first three buckets by strong beam loading, but for subsequent buckets it is close to $2\pi c/\omega_{p}$. The plasma electrons trapped by the plasma wave are color-coded to indicate which acceleration mechanism is at work (see subsequent paragraphs). The accelerated electrons group together in the later plasma buckets, where they gain energy predominantly by interacting with the longitudinal field of the wave associated with SM-LWFA. However, the electrons trapped in the front three buckets of the wake gain net energy predominantly through the DLA process as they interact with the peak-intensity portion of the laser pulse. Relativistic self-focusing helps to maintain the peak intensity of the laser pulse (see dashed green line).

To quantify the contribution of each acceleration mechanism (i.e., SM-LWFA and DLA), we use electron tracking in \textsc{Osiris} to calculate the work done on each electron by the different spatial components of mode~0 (wake) and mode~1 (laser). Separating the fields by mode clearly shows which longitudinal field component is from the plasma wave ($E_{z,m=0}$) and which is predominantly from the evolving laser field ($E_{z,m=1}$). Without separating the fields by mode, effects from the longitudinal laser electric field can be misattributed to wakefield energy gain or loss---for instance, the total work done on some electrons by the $E_{z,m=1}$ field was $-$100~MeV.  This energy loss occurs because the longitudinal component of the laser electric field is roughly $\pi/2$ out of phase with the transverse laser electric field, $E_{x,m=1}$. In fact, the ratio $f$ of average energy gained during betatron oscillation from the parallel ($\left\langle W_\parallel \right\rangle$) and perpendicular ($\left\langle W_\perp \right\rangle$) fields of a Gaussian laser is given~\cite{Pukhov_2002} as
\begin{equation} \label{eq:f}
f \equiv \frac{\left\langle W_\parallel \right\rangle}{\left\langle W_\perp \right\rangle} = -\frac{2c^2\sqrt{2\gamma}}{\omega \omega_p w_0^2},
\end{equation} where $w_0$ is the beam waist and $\gamma$ is the gamma factor of a particle.  The ratio is negative, indicating that an electron in phase with the transverse laser electric field ($E_{x,m=1}$) loses energy due to the longitudinal laser electric field ($E_{z,m=1}$). Prior to this work, the LWFA and DLA processes were differentiated by longitudinal and transverse field components, respectively, rather than separated by mode. Consequently the energy loss from the longitudinal laser electric field was often attributed to SM-LWFA electric fields. Experiments and 2D PIC simulations have shown some contribution of the longitudinal laser field to the acceleration of electrons, but for a near-critical-density plasma using foam targets~\cite{Willingale2018}.

The work done on each electron is then calculated as follows: $W_{\mathrm{LWFA}} = \int\vec{E}_{m=0}\cdot\vec{v}\,dt$ and $W_{\mathrm{DLA}} = \int\vec{E}_{m=1}\cdot\vec{v}\,dt$. We subtract the work done by each mechanism to obtain a relative energy contribution for each electron (see color of tracked electrons in Fig.~\ref{fig:Fig3}), where a positive (negative) value indicates that the net final energy of the electron is mainly coming from DLA (LWFA). DLA is the dominant energy transfer mechanism for electrons trapped in the front two buckets (Fig.~\ref{fig:Fig3}), whereas SM-LWFA dominates in the later buckets of the plasma wave. As mentioned earlier, due to the low-density plasma, a substantial ion channel---where prior results show DLA dominating the acceleration scheme~\cite{Mangles2005}---does not form within the laser fields.

\begin{figure}[ht]
\centering
\includegraphics[trim={0mm 0.8mm 0.4mm 0.1mm}, clip=true, scale = .35]{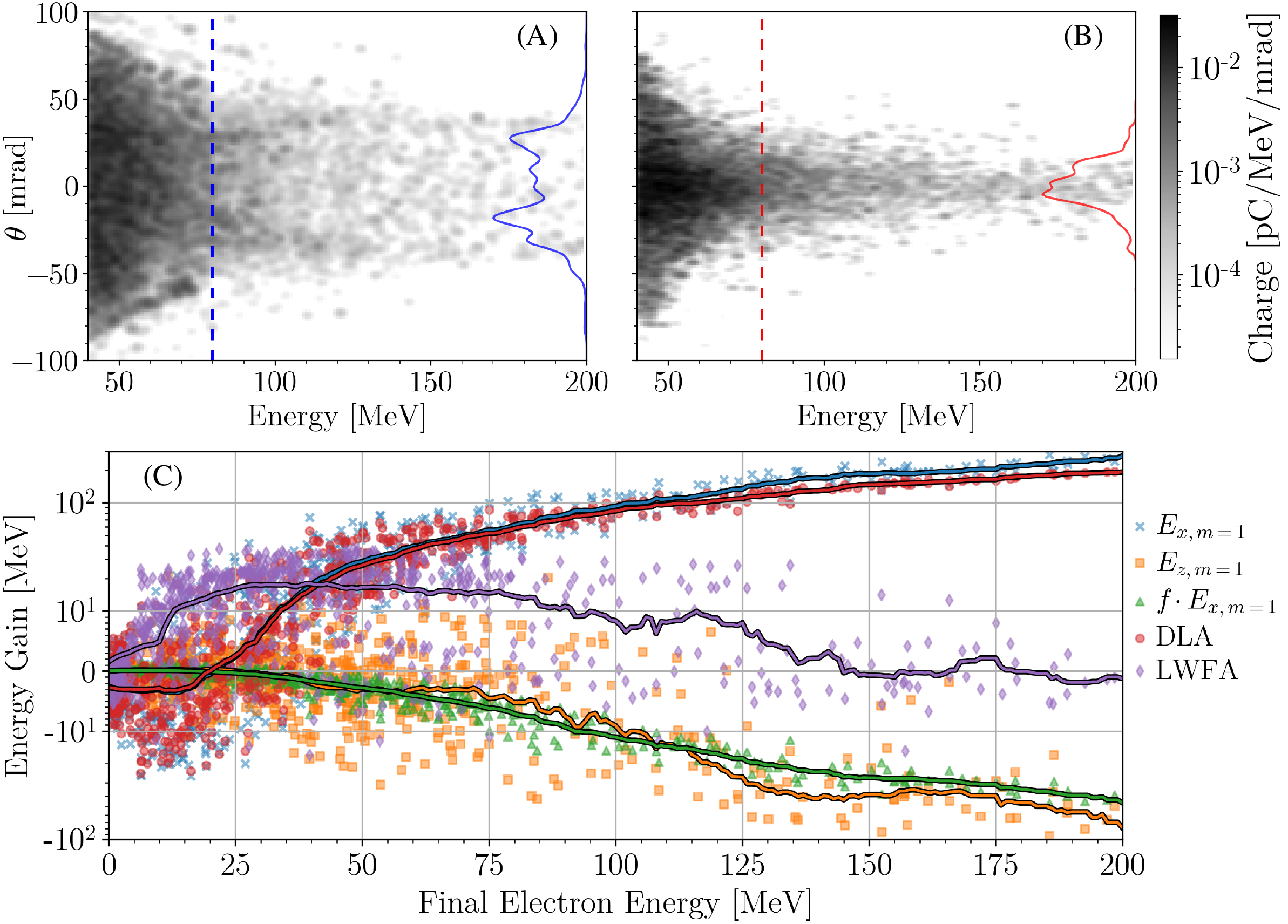}
\caption{Simulated electron spectra dispersed (A)~perpendicular and (B)~parallel to the linear laser polarization direction. A line-out along the dashed line is plotted (solid line) to emphasize the ``horns'' of the dispersed beam.  (C)~The final energy gain due to different field components and mechanisms is shown for numerous tracked electrons (solid lines showing the mean within 20-MeV windows).  All data is shown after 4.68~mm of propagation.}
%(C)~The simulated electron spectrum with two temperature fits given by $A e^{-\frac{E}{T}}$, with amplitude $A$ and temperature $T$ in MeV.
%This energy loss is reasonably predicted with the factor $f$ defined in Eq.~(\ref{eq:f}), for which a focused spot size of $w_0=19\,\upmu$m was assumed. (removed per Félicie comment)
\label{fig:Fig4}
\end{figure}

Figure~\ref{fig:Fig4} clearly shows that DLA dominates the energy gain for higher-energy electrons where the forking is observed.  The accelerated electrons from \textsc{OSIRIS} are dispersed (A)~perpendicular and (B)~parallel to the linear laser polarization using a geometry identical to that shown for the experimental results in Fig.~\ref{fig:Fig2}. In this direct comparison to the experimental data, a forking structure is evident only when the electrons are dispersed perpendicular to the laser polarization. Figure~\ref{fig:Fig4}(C) shows that the longitudinal field, $E_{z,m=0}$, of the self-modulated wake (purple dots and curve) dominates energy gain for electrons with energies up to about 40~MeV, at which point the net DLA  contribution (red dots and curve) becomes comparable. At around 60~MeV, the dominant acceleration mechanism in the simulation shifts to DLA, with the fork becoming visible at about the same energy for the perpendicularly dispersed electrons both in the experiment [Fig.~\ref{fig:Fig2}(A)] and simulation [Fig.~\ref{fig:Fig4}(A)].  The energy loss from the longitudinal laser field [orange dots and curve in Fig.~\ref{fig:Fig4}(C)] is reasonably approximated for many electrons by the calculation $f \cdot E_{x,m=1}$ shown in green (assumed focused spot size of $w_0=19\,\upmu$m), which could be used to estimate the energy loss from the longitudinal laser field in simulations where it is difficult to differentiate between longitudinal wake and laser fields.  Regardless, this energy loss is significant and should not be ignored when considering the energy contribution due to DLA.

The simulated electron spectrum is shown alongside the experimental spectra in Fig.~\ref{fig:Fig1}(B), with $T_1$ and $T_2$ fits of 14.2 and 45.5~MeV, respectively.  Both the number of electrons and the fitted temperatures are similar to the experimental values. The temperature transition, however, occurs near 100~MeV and not 60~MeV (where the DLA mechanism becomes dominant) since the population of electrons accelerated by the DLA mechanism is much smaller than that accelerated by the LWFA mechanism.

In conclusion, we have demonstrated that a picosecond-laser pulse undergoes SM-LWFA in a low-density plasma and that DLA dominates the energy gain of the highest-energy electrons in the absence of a trailing ion channel. This contribution is shown experimentally---and reproduced with PIC simulations---by the forking structure evident in the dispersed electron beam at high electron energies, as well as through the transition between two temperatures in the measured electron spectra at around 100~MeV. This work provides the first direct experimental characterization, confirmed through quasi-3D PIC simulations with mode separation of fields, of DLA in a picosecond, high-energy regime of SM-LWFA, an important result in the development of x-ray sources for HEDS experiments.

This work was performed under the auspices of the U.S. Department of Energy under Contract No. DE-AC52-07NA27344, DE-SC0019010, DE-SC0010064 (UCLA), DE-SC0021057, NNSA grant DE-NA0003873 (UCLA) and NSF grants 1806046, 1804463, 1734315, and 2003354 (UCLA), LLNL subcontract B634451, FA9550-14-1-0045, and supported by the DOE Office of Science Early Career Research Program (Fusion Energy Sciences) under SCW1575-1.

\begin{section}{Supplemental Material}
\begin{subsection}{Experimental Details}

The electron beam was dispersed by a 0.6~T magnet with a large opening aperture (5~cm) perpendicular to the linear laser polarization. A smaller magnet (2.56~cm aperture) with a 1~T field was used to disperse the electrons parallel to the linear laser polarization. The dispersed electron signal passed through three wire fiducials---used to reduce error in the energy calculation~\cite{bib:clayton2010}---and was captured on two BAS-MS image plates separated by 30~cm. The difference in the electron energy spectra in Fig.~1(B) was caused by shot-to-shot variation in the laser energy, pulse width, and quality of the high-power spot size. In an ideal case, these two spectra would be identical.

To determine the energy mapping of the dispersed electron spectra, a two-dimensional code was developed to propagate a beam of electrons with angular spread $\theta$ through a discrete magnetic field (experimentally measured using a hall probe), then map the input electron energy to a location on an image plate. This was done analytically using the following equations:
\begin{equation} \label{eq:magR}
R = \frac{\left[(E+m_e c^2)^{2}-(m_e c^2)^2\right]^{1/2}}{e c B},
\end{equation}
where $R$ is the cyclotron radius, $E$ is the electron kinetic energy, $m_e$ is the electron mass, $c$ is the speed of light, $e$ is the electron charge, and $B$ is the magnetic field amplitude.  The exit angle, $\theta_\textrm{exit}$, at which the electron leaves the discrete magnetic field unit was found as
\begin{equation} \label{eq:thetaexit}
\theta_\textrm{exit} = a \sin\left(\frac{z_m - R \sin \theta_\textrm{in}}{R}\right),
\end{equation}
where $\theta_\textrm{in}$ is the angle at which the electron entered the magnetic field and $z_m$ is the length of the magnetic field element. The horizontal shift from center, $\Delta x$, at which the electron leaves the magnetic field unit was calculated as
\begin{equation} \label{eq:xdisp}
\Delta x = R \left(\cos\theta_\textrm{in}-\cos\theta_\textrm{exit}\right).
\end{equation}
Summing all of the $\Delta x$ contributions from each magnetic field element shows where the electron leaves the back of the magnet, given an initial input displacement and angle. From the output displacement and angle, a line can be drawn through the two detector planes and matched with the wire fiducial shadow locations on the image plates. Additionally, each image plate in the code is able to translate laterally to account for human error during the installation of each image plate during the experiment. By allowing the image plates to laterally translate results in an energy mapping solution for each image plate location. To account for this, any solution where electrons are unable to pass through all three wire fiducials are discarded. This was calculated for electrons with energies ranging from 1 to 1000~MeV and an angular spread of 2~rad FWHM. Following the two-screen wire fiducial method, we are able to bound the angular spread of the input electron beam and reduce the error in the energy mapping. The error shown in figure 1 is calculated using a combination of electron beam angular spread and uncertainty in image plate lateral location. However, due to space limitations in the target chamber, the experimental data taken using the smaller 1~T magnet did not have the space for a second image plate, resulting in the larger error seen in Fig.~1(B).
\end{subsection}

\begin{subsection}{PIC Simulations}

The quasi-3D algorithm uses fields and currents defined on an $r$-$z$ grid and expanded in azimuthal modes; to simulate LWFA we used modes 0 and 1, where mode 0 (1) mainly captures the wake (laser) fields.  In addition, we used a customized field solver that corrects both for dispersion errors of light in vacuum and the time-staggering error of the magnetic field in the Lorentz force~\cite{Miller2020}.  The simulations were carried out in the speed-of-light frame (moving window) with a box of size $95 \times 23.6\,c/\omega_p = 854 \times 212\,\upmu$m (the second dimension corresponding only to a half-slice, $r$ starting at 0), where $c/\omega_p = 8.991\,\upmu$m for a density of $3.5\times 10^{17}\,\mathrm{cm}^{-3}$.  The number of grid points used was $48000 \times 256 = 1.2 \times 10^7$, with a time step of $\Delta t=5 \times 10^{-4} \omega_p^{-1} = 30$~fs.  The laser pulse had an amplitude of $a_0=2.7$, intensity FWHM of 1~ps, spot size of $w_0=25.5\,\upmu$m and Rayleigh length of $z_R = 1.94$~mm.  We used a preformed plasma with a density upramp of $500\,\upmu$m followed by a constant-density region, with the laser focused halfway through the upramp.  Mobile ions were included along with electrons, with each species having 4 particles per $r$-$z$ cell and 8 particles in the $\theta$ direction, making for a total of $7.9 \times 10^8$ particles.

Though the experimental gas jet was 10~mm in length, we found electron energies comparable to those from the experiment after a propagation distance of only 4.68~mm in the simulation.  This discrepancy is likely caused by the non-ideal laser spot used in the experiment, while the simulation used an ideal Gaussian spatial profile at the laser focus.  The non-ideal laser spot could necessitate additional time to form an SM-LWFA and begin trapping electrons in the experiment, whereas this happened over a shorter distance in the simulation.
\end{subsection}
\end{section}

\bibliography{Tese_PhD}
\end{document}